\documentclass[%
 reprint,
superscriptaddress,
 amsmath,amssymb,
 aps,
]{revtex4-1}

\usepackage{graphicx}
\usepackage{dcolumn}
\usepackage{bm}
\usepackage{color}
\usepackage{tikz}
\usepackage{pgfplots}
\pgfplotsset{every tick label/.append style={font=\tiny}}

\begin{document}

\preprint{APS/123-QED}

\title{Spatial optical mode demultiplexing as a practical tool for optimal transverse distance estimation}

\author{Pauline Boucher}
\affiliation{Laboratoire Kastler Brossel, UPMC-Sorbonne Universit\'es, ENS-PSL Research University, Coll\`ege de France, CNRS; 4 place Jussieu, F-75252 Paris, France}
\affiliation{Cailabs, 38 boulevard Albert 1er, 35200 Rennes, France}
\author{Claude Fabre}
\affiliation{Laboratoire Kastler Brossel, UPMC-Sorbonne Universit\'es, ENS-PSL Research University, Coll\`ege de France, CNRS; 4 place Jussieu, F-75252 Paris, France}
\author{Guillaume Labroille}
\affiliation{Cailabs, 38 boulevard Albert 1er, 35200 Rennes, France}
\author{Nicolas Treps}
\email{nicolas.treps@upmc.fr}
\affiliation{Laboratoire Kastler Brossel, UPMC-Sorbonne Universit\'es, ENS-PSL Research University, Coll\`ege de France, CNRS; 4 place Jussieu, F-75252 Paris, France}

%
%
%


\date{\today}

\begin{abstract}

We present the experimental implementation of simultaneous spatial multimode demultiplexing as a distance measurement tool. We first show a simple and intuitive derivation of the Fisher information in the presence of Poissonian noise. We then estimate the distance between two incoherent beams in both directions of the transverse plane, and find a perfect accordance with theoretical prediction, given a proper calibration of the demultiplexer. We find that, even though sensitivity is limited by the cross-talks between channels, we can perform measurements in 2 dimensions much beyond Rayleigh limit with a large dynamic.

\end{abstract}

\maketitle

\section{\label{sec:level1}Introduction}

The measure of the distance between two incoherent point sources is a well-known problem since it amounts to determining the resolution of an optical imaging system. In the general case, this resolution is given by the characteristic width of the point spread function of the system, which is proportional to its numerical aperture. Early resolution criteria derived in different fields --- such as the Abbe \cite{Abbe1873}, Rayleigh \cite{Rayleigh79} or Sparrow \cite{Sparrow1916} criteria --- give slightly different definitions for the resolution power of a given system. As pointed out in \cite{Goodman2000}, the sole information of the point spread function is in general not sufficient for the thorough characterisation of the resolution power and the signal to noise ratio plays a significant role in its determination. In the last decades, fields such as astronomy and microscopy have driven the efforts to access better optical resolutions --- with the emergence of techniques such as aperture synthesis or super-resolved fluorescence microscopy \cite{Moerner89, Hell94, Klar8206, Betzig1642}. Analytical continuation \cite{ToraldodiFrancia:69, Kolobov00} is another approach which has also been studied in the context of radio astronomy \cite{Lo61} and electronic information \cite{WOLTER1961155}.   

In a general imaging context, the problem of the measurement of the distance between two beams is expressed in terms which are very similar to those describing the measurement of the position of a single beam. In the latter case, sensitivity reaching the Cram\'er-Rao bound can be attained for intense beams using homodyne detection, with coherent states or squeezed states \cite{Hsu04, Delaubert06, Delaubert06.1}. These techniques perform field measurements on a spatial mode termed the \textit{detection mode}. Using similar techniques, other parameters of a beam can be measured \cite{Pinel12}. However, the use of field measurements results in the fact that these schemes are limited to measurements of a single optical source (or two coherent sources) and cannot be applied directly to the measurement of the distance between two incoherent sources. 

To our knowledge, the first mention of the use of spatial mode demultiplexing for the problem of distance measurement was made in \cite{Helstrom73}. In this work, the physical situation studied is similar to the one we are interested in (two incoherent point sources), but the problem adressed is that of hypothesis testing. A recent surge of interest was sparked in the last years regarding the quantum Cram\'er-Rao bound on the precision of the measurement of the separation between two incoherent, diffraction-limited, low-intensity sources \cite{Tsang16, Tsang16.1, Nair16, Lupo16}. The simultaneous measurement of other parameters has been investigated in \cite{Rehacek17, Rehacek17.1, Napoli19}. The link between position determination and higher-order modes was also highlighted in a number of different contexts \cite{Puppe_2004}.

In \cite{Tsang16.1, Lu16, Tsang17, Ang17}, spatial mode demultiplexing coupled with intensity measurement is introduced as a measurement scheme which saturates the quantum Cram\'er-Rao bound. If the point-spread function of the optical system is approximated by a Gaussian, the optimum mode basis for demultiplexing is the Hermite-Gauss mode basis. In \cite{Tsang16.1}, it is shown that in the limit of an infinite number of measurement modes, the information gathered is equal to the Fisher information for any value of the separation between the sources. The single-mode version of this measurement scheme, which is concerned with the regime of very small separations, was experimentally tested in recent years \cite{Paur16, Tang16, Yang16, Tham17}. However, the general case of multimode demultiplexing had not yet been investigated.

\begin{figure}[b]
  \centering
  \includegraphics[width=0.38\textwidth]{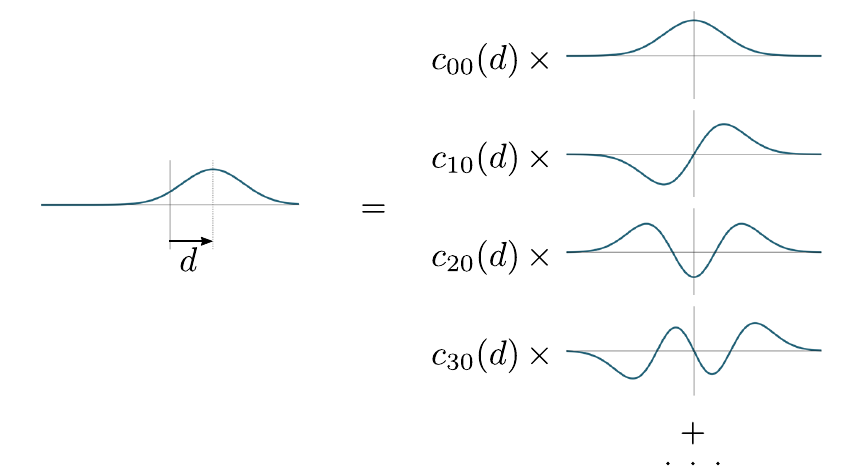}
  \caption{A Gaussian beam displaced in the transverse direction can be described using its decomposition on the Hermite-Gauss mode basis defined using the reference position which is used to define the displacement. This mode basis does not depend on the value of the displacement $d$.}
  \label{fig:decomposition}
\end{figure}

In this work, we measure the distance between two incoherent point sources using multimode spatial demultiplexing of nine modes, implementing the measurement scheme described in \cite{Tsang16}. We demonstrate that in the case of small displacement, the derivations of the semi-classical and quantum Cram\'er-Rao bounds give the same result. Using a Multi-Plane Light Conversion (MPLC) system \cite{Morizur:10, Labroille:14} we perform intensity measurements on the projections of the electric field on the Hermite-Gauss mode basis. The demultiplexing of more than one mode allows to measure the distance between two sources even outside the range of very small displacement, and also enables the simultaneous measurement of different parameters such as, in this work, the projection of the separation between the two sources on both directions of the transverse plane.

\section{\label{sec:level1}Theoretical framework}

In this section we propose a simple and intuitive derivation of the Cram\'er-Rao bound in the presence of Poissonian noise. For the measurement of the separation of two incoherent, diffraction-limited sources, with the important assumption that the centroid of the two points is known, the quantum Fisher information was calculated in \cite{Tsang16.1, Nair16, Lupo16}, for very low intensity beams. In practice, each beam is approximated by a Gaussian beam of waist $w_0$. $N$, the total number of photons during one coherence time $\Delta \tau$, is taken to be small compared to 1. The result of \cite{Tsang16.1} can then be expressed as \footnote{In \cite{Tsang16.1, Nair16, Lupo16}, the quantum Fisher information $I_F^q$ is expressed in terms of $\Delta k^2$ the spatial variance of the derivative of the point-spread function. Under the approximation we make, namely that the point-spread function is a Gaussian beam, we have $\Delta k^2 = 1/w_0^2$}:
\begin{equation}
I_F^q \simeq \frac{N}{w_0^2}.
\end{equation}

We show here that a semi-classical approach gives the same result in a very simple and intuitive way. Let us first derive the semi-classical Fisher information for intensity measurements of a single light beam which depends on the displacement $d$ along the transverse direction $\mathbf{e}$. We consider the mean value of the electric field of one individual beam $E(\mathbf{r}; d) = 2 \sqrt{N}\mathcal{E}u_0\left(\mathbf{r}; d \right)$ with $u_0\left(\mathbf{r}; d \right)$ the normalized mean electric field, $\mathbf{r}=(x,y)$ the transverse coordinate and $\mathcal{E}=\sqrt{\hbar \omega/2\epsilon_0 cT}$ ($\omega$ is the optical angular frequency, $c$ the speed of light and $T$ the integration time of the detectors used in the detection scheme). We define the \textit{detection mode} as the normalized derivative of $u_0$ with respect to the displacement:
\begin{equation}
w(\mathbf{r}) = \left.\frac{\partial u_0\left( \mathbf{r}; d \right)/\partial d}{\left|\left|\partial u_0\left( \mathbf{r}; d \right)/\partial d\right|\right|}\right|_{d=0}.
\end{equation}
We assume that the total intensity of the beam is independent of the parameter $d$ --- which is indeed the case. One can then easily calculate the classical Fisher information in the presence of Poissonian noise, and we find~\cite{Delaubert08} 
\begin{equation}
I^c_F = 4N\int{\left[\left. \frac{\partial |u_0\left( \mathbf{r}; d \right)|}{\partial d}\right|_{d=0}\right]^2} d\mathbf{r}.
\end{equation}

Let us consider now the problem of estimating the distance between two incoherent beams. It can be deduced from the single beam case by describing the source as the incoherent sum of two electric fields with opposite displacements. In the case of Gaussian beams, so that $u_0\left( \mathbf{r}; d=0\right)$ is equal to the zero-order Hermite-Gauss mode $HG_{00}\left( \mathbf{r}\right)$, 
$E_{i\in\{1,2\}} = 2  \sqrt{\alpha_i N} HG_{00} \left(\mathbf{r}+ \gamma_i \frac{d}{2}\mathbf{e}\right)$ with $\alpha_1 + \alpha_2 = 1$, $\gamma_i = \pm 1$, $d$ is the total distance between the sources and $\mathbf{e}$ the unitary vector of the separation direction. In this situation, the detection mode is the first Hermite-Gauss mode in the direction defined by $\mathbf{e}$: $w(\mathbf{r})=HG_{10}(\mathbf{r})$.

The Fisher information associated with one source is $I^c_{F,i} = \alpha_i N/w_0^2$, still assuming Poissonian noise. 
Furthermore, as both sources are independent, we can consider that the random variables consisting of the number of photons hitting the detector originating from each of the sources are independent --- and the information yielded by two independent random variables is the sum of the information from each random variable. As a result, the classical Fisher information of the complete system can be written as
\begin{equation}
I^c_F = I^c_{F,1} + I^c_{F,2} = \frac{N}{w_0^2}.
\end{equation}
Hence, for this specific problem, we also have $I^c_F = I^q_F$. Remarkably, in order to reach the corresponding Cram\'er-Rao bound for each of these beams, one  has to measure the amplitude quadrature of the detection mode, and, because the beams are identical, the detection mode is the same for both beams. So, measuring this detection mode leads to quantum Cram\'er-Rao limited distance estimation, as the intensity of the sum is the sum of the intensities in the incoherent case. One should insist on the fact that in the considered regime this result is obvious. Indeed, it is well known that measuring the amplitude of the detection mode for a single beam leads to quantum Cram\'er-Rao limited sensitivity and that the distance between two incoherent beams is the sum of two symmetric displacements.
We also recall that the analysis presented in this part is valid in the small displacement regime. In the experimental work presented in the next part, we depart from this regime and study the case of arbitrary displacements.

\section{\label{sec:level1}Experimental work}
\begin{figure}[htb]
  \centering
  \includegraphics[width=0.49\textwidth]{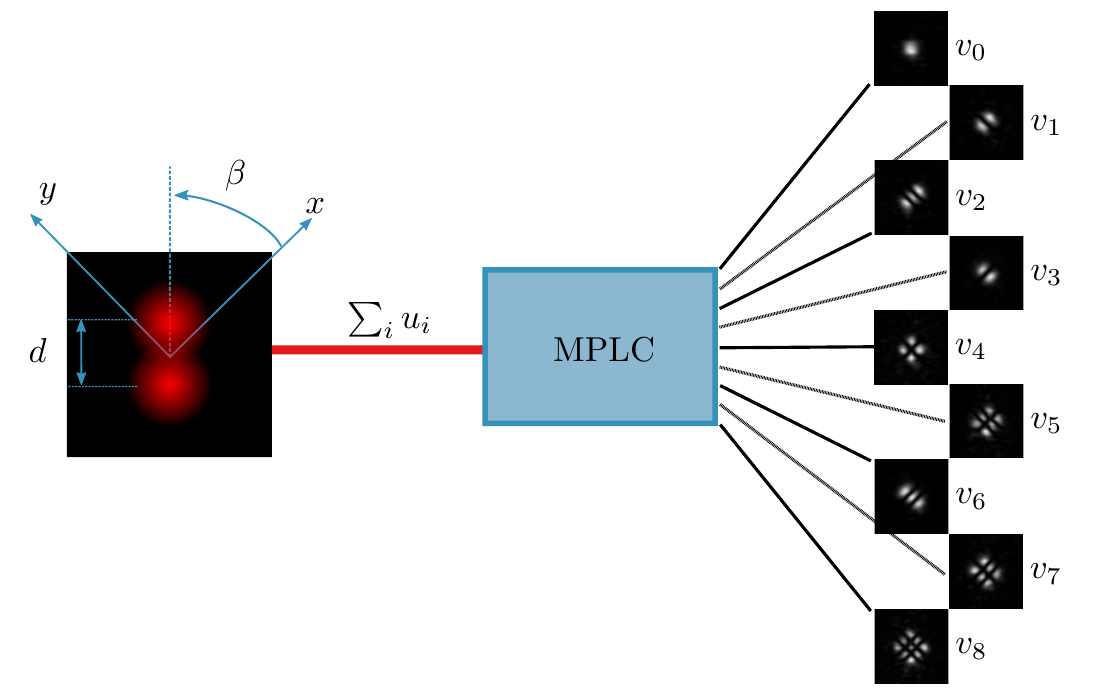}
  \caption{The beams which enters the MPLC system can be described as a superposition of modes $u_{0\leq i\leq 8}$. Each input mode is demultiplexed and sent to a spatially separated spatial mode $v_{0\leq i \leq 8}$. Each of these modes is coupled into a separate single-mode fiber.}
  \label{fig:experimentalscheme}
\end{figure}

We developed an experimental setup to assess the performances of a demultiplexing system on the measurement of the distance between two incoherent sources. We still make the assumption that the position of the centroid of the two sources is already known. The key element of this setup is a MPLC system (Cailabs PROTEUS-C) which allows to perform intensity measurements on several modes of the Hermite-Gauss basis, whose orientation define the $\mathbf{x}$ and $\mathbf{y}$ vectors of the transverse plane. We introduce the angle $\beta$ between vectors $\mathbf{d}$ and $\mathbf{x}$. The two fields can be expressed as
\begin{equation}
E_i\left(\mathbf{r};\gamma_i \mathbf{d}/2\right) = \sum_{n,m} c_{n,m}\left(\gamma_i \mathbf{d}/2 \right) HG_{nm}\left(\mathbf{r}\right)
\end{equation}
with $\mathbf{d} = \left(d \cos{\beta}, d \sin{\beta}\right)$ (see figures \ref{fig:decomposition} and \ref{fig:experimentalscheme}).

\begin{figure}[b]
\centering
\includegraphics[width=0.38\textwidth]{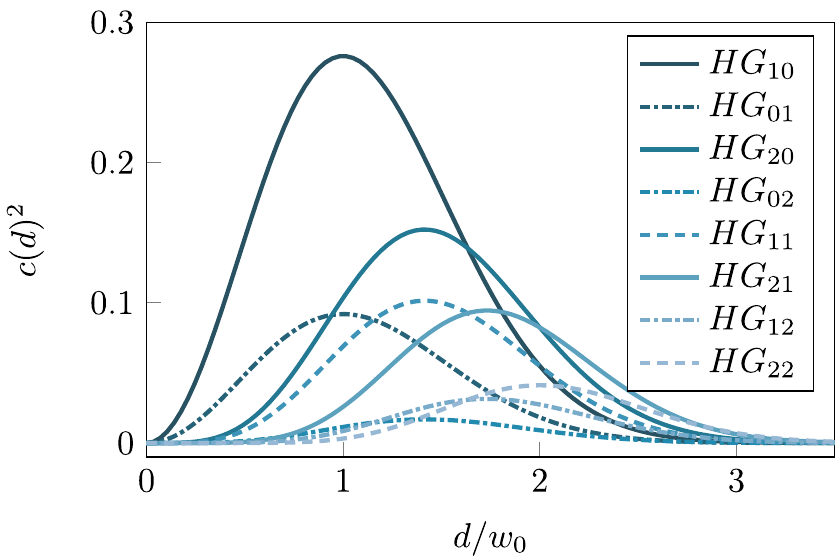}
\caption{Theoretical projections on the Hermite-Gauss mode basis for a displaced Gaussian beam as a function of the displacement $d$ with $\beta = \pi/6$.}
\label{fig:projectionsPis6}
\end{figure}

Figure \ref{fig:projectionsPis6} displays an example of the theoretical coefficients $c_{nm}^2$ as a function of $d$ for $\beta = \pi/6$. It illlustrates the fact that displacements in both directions of the transverse plane can be measured. Using the MPLC system allows to directly and simultaneously measure these coefficients.

\subsection{\label{sec:level2}MPLC design and characterisation}

The MPLC system is a spatial mode multiplexing tool which can in theory implement any kind of spatial mode basis change on a finite number of modes. The MPLC system we use is a 9-mode demultiplexer. The input modes are free-space co-propagating Hermite-Gauss modes ($u_{0\leq k \leq 8} \simeq HG_{0\leq n,m \leq 2}$) with a waist $w_0 = 227 \mu$m. The output modes are spatially separated single mode Gaussian beams which are coupled into single mode fibers ($v_{0\leq k \leq 8}$). This system thus allows to perform intensity measurements on 9 modes of the Hermite-Gauss basis simultaneously. Although there is no theoretical limitation to the quality of the mode basis change a MPLC can implement, a fabricated MPLC is never perfect: the modes which are demultiplexed are slightly different from the theoretical modes (in this case the Hermite-Gauss modes). One important measure of the MPLC's performances is the measure of the cross-talk between channels. To measure these quantities, we inject a free-space $HG_{00}$ beam in the system and measure the intensity in all the different output channels. The optical power measured in the mode $HG_{nm}$ is $P_{nm}$ and the cross-talk of mode $(n,m)$ with the $(0,0)$ mode is given by $r^{XT}_{nm}=10*\log_{10}\left(P_{nm}/P_{00}\right)$. Table \ref{tab:XT} presents both the specified cross-talk values and the experimentally measured values. 
\begin{table}[b]
	\caption{Specified (spe) and experimental (exp) values of the crosstalk values from mode $HG_{00}$ into the higher order $HG_{nm}$ modes.}
	\label{tab:XT}
\begin{ruledtabular}
	\centering
\begin{tabular*}{\linewidth}{@{\extracolsep{\fill}}ccccccccc}
		 dB & $r^{XT}_{10}$  & $r^{XT}_{01}$  & $r^{XT}_{20}$  & $r^{XT}_{02}$  & $r^{XT}_{11}$  & $r^{XT}_{21}$  & $r^{XT}_{12}$  & $r^{XT}_{22}$ \\
		\colrule
		spe & -26 & -28 & -27 & -26 & -45 & -31 & -38 & -24 \\
		exp & -26 & -26 & -26 & -23 & -38 & -31 & -38 & -28 \\
\end{tabular*}
\end{ruledtabular}
\end{table}
Another important difference with the ideal case is that the different ``channels" of the MPLC do not have identical losses. These channel-specific losses can be evaluated when one possesses two identical systems by measuring them in a ``back-to-back" configuration \footnote{The two identical MPLC are aligned so that the free-space output of one system corresponds to the free-space input of the second one. Light is then injected into a single-mode fiber of the first system and the intensity at the corresponding fiber of the second system is measured. This measurement scheme allows to evaluate the channel-specific losses}. In our case, we chose to measure the losses by injecting light into the different single-mode fibers and measuring the intensity exiting the MPLC system. 
\begin{table}[b]
	\caption{Intrinsic efficiency coefficients of the MPLC system relative to mode $HG_{00}$ efficiency.}
	\label{tab:intrinsicgain}
\begin{ruledtabular}
	\centering
\begin{tabular*}{\linewidth}{@{\extracolsep{\fill}}ccccccccc}
		  $\eta_{00}$  & $\eta_{10}$  & $\eta_{01}$  & $\eta_{20}$  & $\eta_{02}$  & $\eta_{11}$ & $\eta_{21}$ & $\eta_{12}$ & $\eta_{22}$ \\
		\colrule
		 1.0 & 0.9& 1.0 & 0.9 & 1.1 & 0.7 & 0.8 & 1.0 & 0.7\\
\end{tabular*}
\end{ruledtabular}
\end{table}
Table \ref{tab:intrinsicgain} gives the measured intrinsic efficiency of the MPLC system. It should be noted that this specific measurement scheme does not discriminate the cross-talk from the losses we wish to measure.

\subsection{\label{sec:level2}Assessment of the role of cross-talk in the sensitivity of the setup}

In the context of our experiment, the cross-talk values of table \ref{tab:XT} represent an offset for the measurement of the displacement (see \ref{sec:annexB} for detailed derivation). Indeed, the intensity measured in the first-order mode is proportional to $\left(d/w_0 + p^{01}_{00}\right)^2$ with $d$ the displacement and $p^{01}_{00} = \sqrt{P_{01}/P_{00}}$. We suppose, as is the case in this experiment, that the minimum crosstalk intensities are greater than the sensitivity of the power-meter we use. With the cross-talk values of the theoretical system used in the experiment, we can write:
\begin{align*}
d^{\text{offset}}_{10}/w_0&=\sqrt{p^{10}_{00}} = \sqrt{10^{-26.29/10}} = 4.9 \,10^{-2}\\
d^{\text{offset}}_{01}/w_0&=\sqrt{p^{01}_{00}} = \sqrt{10^{-27.45/10}} = 4.2 \, 10^{-2}.
\end{align*}

The phenomena of cross-talk is not specific to the MPLC tool and plays a role in any demultiplexing system. Indeed, it finds its origin both in technical limitations and fabrication errors as well as errors or approximation made in the determination of the spatial profile of the signal beam (for instance the point-spread function of the optical system in the diffraction-limited case). Though small, these errors always introduce non-zero projections on the higher-order modes of the chosen basis meaning that at high power, the precision of a demultiplexing system will always be limited by the cross-talk. The characterization of such quantities is thus an essential step of the assessment of any kind of demultiplexing system. In the present paper, we consider the worst case scenario where cross-talks is a noise of unknown source setting a limit for the sensitivity. However, one should note that is is possible to derive the Fisher information in the presence of cross-talk, given that the corresponding basis change is perfectly known, which leads different sensitivity scaling but would, upon the utilisation of the optimal estimator, lead to an improved sensitivity compared to what we estimate here \cite{Gessner:2020vx}.

\subsection{\label{sec:level2}Experimental scheme}

In our experimental scheme, the output of a fibered SLD (Thorlabs S5FC1005S, 50 nm bandwidth, $\lambda = 1550$ nm, output power 22 mW) is launched into a Mach-Zender-like setup. A mirror mounted on a micro-meter translation stage before the first beam-splitter of the setup allows to translate the beam with respect to the optical axis. Two Dove prisms aligned at $\pi/4$ and $-\pi/4$ with respect to the plane of the optical table allow to produce symmetrical displacement for the beams in both arms of the setup. The coherence length of the source is $L = c/\left(\pi \Delta \nu\right) = 15 \mu$m. The length of each arm of the Mach-Zehnder-like setup is of the order of 30 cm. After going through the setup, no interference could be observed between the two beams, which confirms that they are indeed incoherent. The two beams are injected into the MPLC system using a telescope which scales the waist of the beams in order to match the designed waist size of the MPLC system. This configuration allowed to perform both single and double beam displacement measurements.

\subsection{\label{sec:level2}Experimental results}
\begin{figure}[htbp]
\centering
\includegraphics[width=0.5\textwidth]{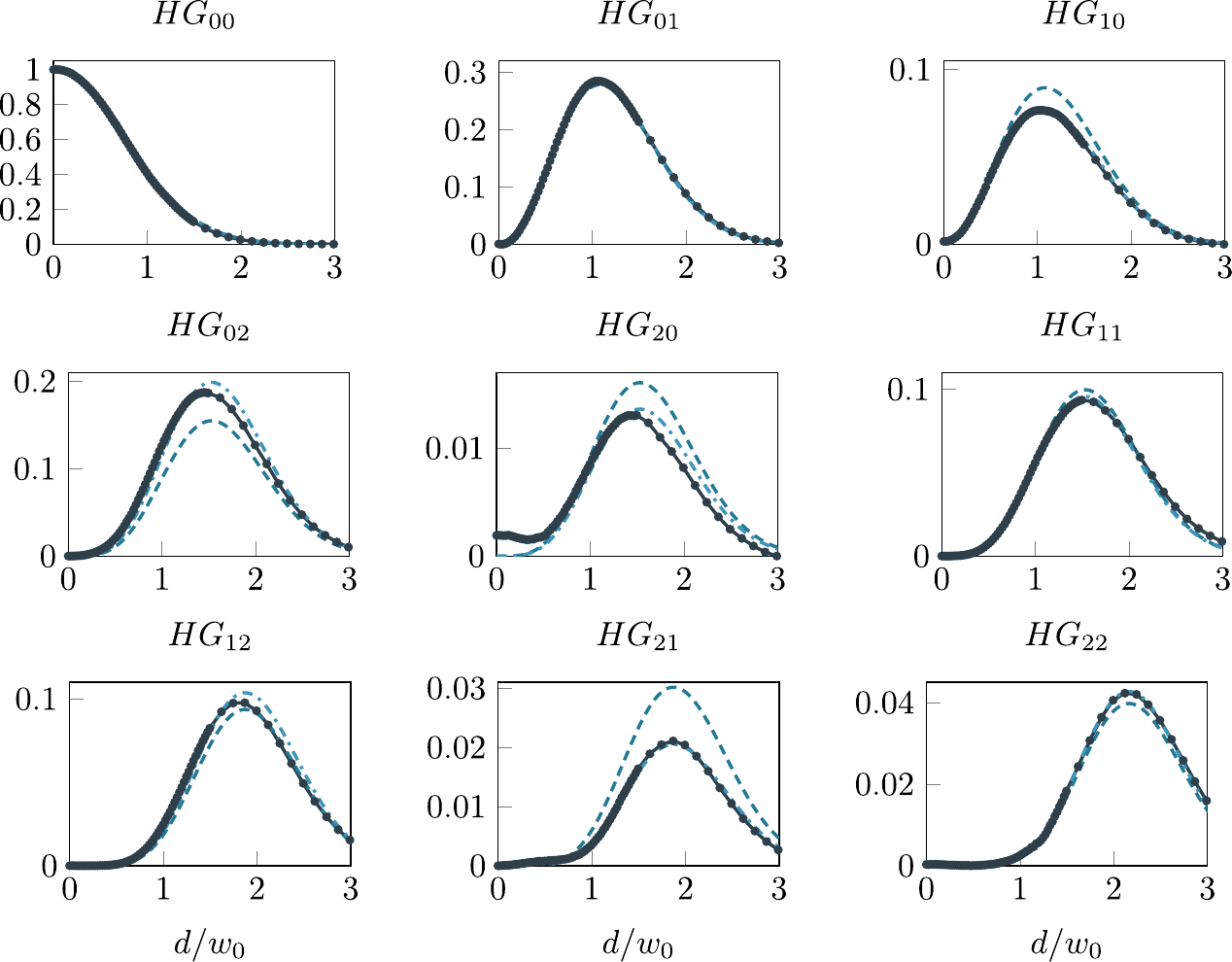}
\caption{Normalized intensity measured on each of the MPLC output (solid line). In dashed line is the theoretical model for the extracted parameter of displacement and angle. In dashed dotted, we fit the data with an additional gain which depends on the channel.}
\label{fig:1207fitA}
\end{figure}

Each measurement run is performed by introducing displacement in steps using the translation stage. For each step, we record the intensity of all the output modes. Figure \ref{fig:1207fitA} plots the intensity profiles of all the output modes normalized to the intensity of the $HG_{00}$ mode as the distance between the two beams is scanned from 0 to 3 $w_0$. Corresponding theoretical plots of the intensity profiles are also shown, normalized using the gain coefficients of table \ref{tab:intrinsicgain}.

We first observe that all the plots correspond generally with what is expected theoretically, in particular for the low displacement regime, or for the first order modes where there is a perfect agreement between theory and experiment. This demonstrates that the MPLC is a highly valuable and compact tool for distance measurement. However, we see that for larger displacement (i.e. corresponding to more than one beam waist) discrepancies appear between the theoretical model and the experimental data. We note furthermore that this discrepancy appear mainly for modes whose output is the weakest, i.e. the modes contributing the least to the evaluation of the measurement. Also, we note that the main effect is a global gain effect, but that each curve preserves a shape that corresponds to the theoretical one. We thus chose to perform a fit of the data using free gain coefficients. Table \ref{tab:fittedgain} displays these coefficients and the corresponding plots are presented on figure \ref{fig:1207fitA}.
\begin{table}[!b]
	\caption{Fitted gain parameters.}
	\label{tab:fittedgain}
\begin{ruledtabular}
	\centering
\begin{tabular*}{\linewidth}{@{\extracolsep{\fill}}ccccccccc}
		$g_{00}$  & $g_{10}$  & $g_{01}$  & $g_{20}$  & $g_{02}$  & $g_{11}$ & $g_{21}$ & $g_{12}$ & $g_{22}$ \\
		\colrule
		1.0 & 1.0 & 0.9 & 1.3 & 0.9 & 1.0 & 1.1 & 0.7 & 1.1\\
\end{tabular*}
\end{ruledtabular}
\end{table}
The final demultiplexing results corresponds perfectly to the theoretical model and demonstrates the potential of the MPLC system for the determination of the distance between two incoherent sources, given a proper calibration procedure. We note that the assessment of the channel-dependent gain coefficient is a key element for a MPLC system to be used as a precise measurement instrument. Furthermore, the same measurements were performed for a single source (by blocking one of the arms of the interferometer) which showed similarly good results.

\begin{figure}[b]
  \centering
  \includegraphics[width=0.48\textwidth]{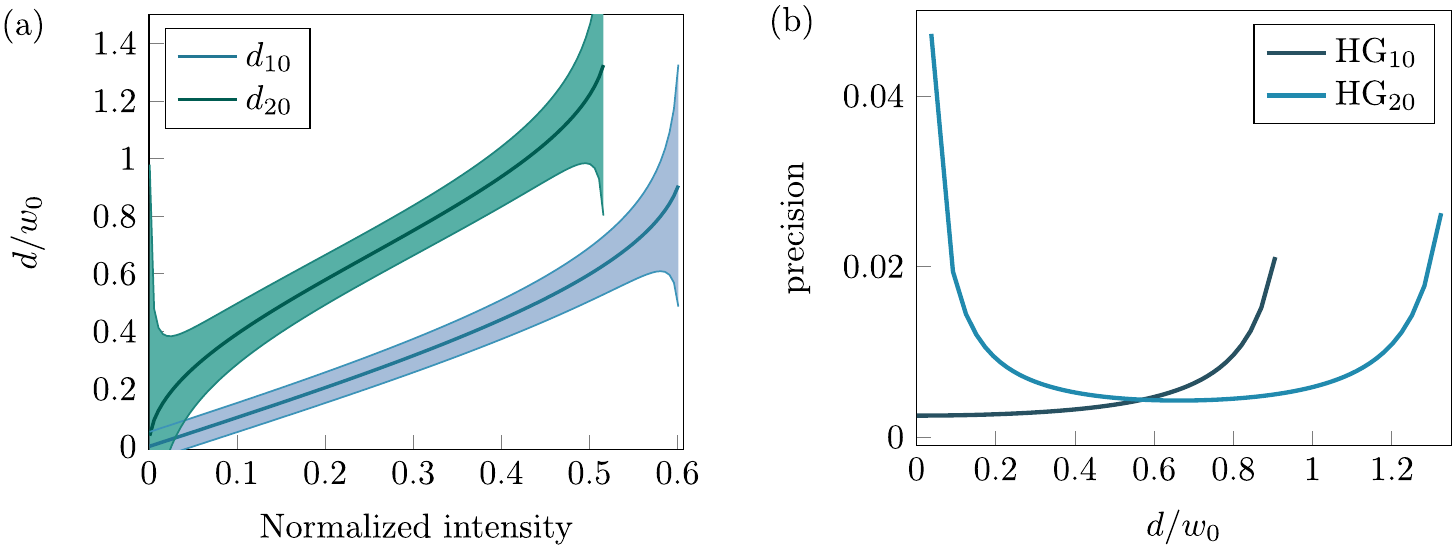}
  \caption{(a) Displacement as a function of the measured intensity in the modes HG$_{01}$ and HG$_{02}$: $d_{0i}(I)$. The error bars correspond to $\pm xt_{01} \times \partial d_{0i}/ \partial I$ with $I$ the normalized intensity. For clarity $\pm xt_{01}$ is taken equal to 0.1. (b) Precision as a function of the source separation for measurement using modes HG$_{01}$ and HG$_{02}$. For this figure,  $\pm xt_{01}$ value is the one measure experimentally in Table I.}
  \label{fig:sensitivity}
\end{figure}

Finally, we can now use the MPLC to measure source separation and consider the sensitivity of the measurement imposed by the cross-talks. Given the performed calibration, the inferred distance upon intensity measurement is simply obtained inverting the curves displayed in  \ref{fig:1207fitA}. We plot in figure \ref{fig:sensitivity}a the distance versus measured intensity on modes HG$_{01}$ and HG$_{02}$, and introduce error bars induced by cross-talks (i.e. given by the derivative of the curve times the amount of cross-talks). In order to appreciate the relative effect more clearly we considered a cross-talk equal to $10^{-1}$. We see, as expected, that for HG$_{01}$ precision remains constants from small displacement to $d/w_0\approx 0.5$ and then diverges, while for HG$_{02}$ it diverges for small displacement. In figure  \ref{fig:sensitivity}b we thus plot the precision of the inferred distance from measurement either using HG$_{01}$ or HG$_{02}$, respectively, now using the experimentally measured cross-talks. We see that, depending on the displacement, one should use either one or the other output (or a combination of both to obtained an optimal estimator, which we did not plot in the figure). Importantly, cross-talks imposed a sensitivity about $2.10^{-3}$ for a measurement of $d/w_0$ on a broad range of values, up to about $d/w_0\approx 1.2$. We thus demonstrate  highly sensitive measurement with a very large dynamic range. Even though we are still far from the Cram\'er Rao bound because of the cross-talk, which could be improved using a more complex optimal estimator \cite{Gessner:2020vx}, our apparatus displays sensitivity going much beyond the Rayleigh limit, in two transverse dimensions.

\section{\label{sec:level1}Conclusion}

In this work, we make the demonstration of the use of spatial mode demultiplexing as a measurement tool of the distance between incoherent sources. In a first part, we show that a simple semi-classical derivation, in the small displacement regime, reaches the same qualitative and quantitative conclusion as a full quantum derivation, thereby giving an intuitive insight to the use of intensity measurements and multiplexing for parameter estimation. We then present an experimental implementation of this protocol, which is conducted in the high intensity and large displacement regime. The latter shows a very good agreement with theory and validates its use. In doing so, we also highlight the role of the cross-talk between the different modes of the demultiplexing system as an important and unavoidable element to take into account in the performance evaluation of such systems. Finally, we note that this measurement scheme can also be adapted to the measurement of other spatial parameters by tailoring the demultiplexing mode basis to the parameter of interest.

\begin{acknowledgments}
The authors wish to thank O. Pinel for discussions and constructive comments. C. Fabre and N. Treps are members of the Institut Universitaire de France.
\end{acknowledgments}
\pagebreak
\appendix

\section{Alignment of the setup}
\label{sec:annexA}
The alignment of the MPLC system with the rest of the experimental setup is a critical point. Indeed, we wish to measure a displacement with respect to a reference position: this reference position is given by the modes for which the MPLC is designed. The waist size and position (in the longitudinal direction) of the input beam must be carefully matched, and the beam must also perfectly aligned, both in position and direction in the transverse plane. Since the system we use is precisely designed to measure a misalignment, we use it to align our setup. A coarse alignement is made by maximizing the intensity measured on the $HG_{00}$ mode. In a second step, we measure the cross-talk values and minimize them. The quality of the alignment is assessed by how close to the theoretical cross-talk values the measured cross-talk values are.

\section{Sensitivity of apparatus}
\label{sec:annexB}
\begin{figure}[ht]
  \centering
  \includegraphics[width=0.4\textwidth]{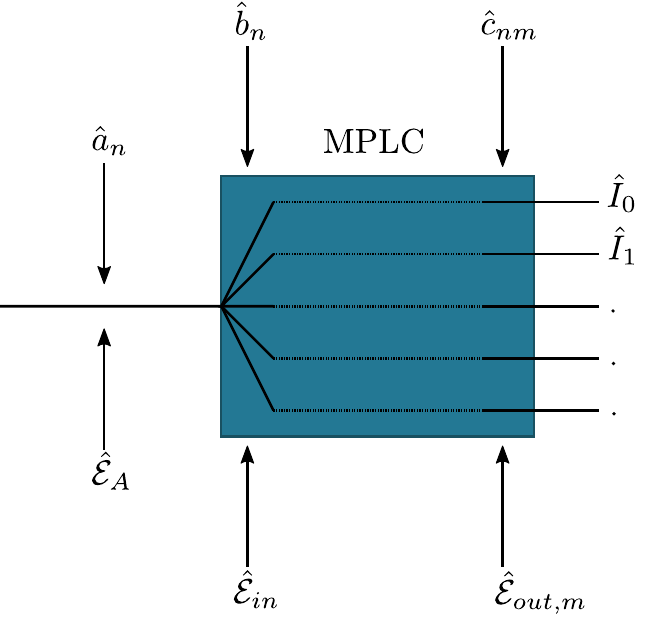}
  \caption{We introduce in this figure the different notations used to describe the successive steps of the demultiplexing process. The complex field operators are denoted as $\hat{\mathcal{E}}$ and their associated annihilation operators as $\hat{a}, \hat{b}, ...$. The intensity operators at the outputs of the demultiplexor are written as $\hat{I}$.}
  \label{fig:defmodes}
\end{figure}

We make here a derivation of the sensitivity one can reach using a demultiplexing system, in the case of a single beam in a coherent state. All notations are defined on figure \ref{fig:defmodes}.
We express the complex field operator of a single beam as
\begin{equation}
\hat{\mathcal{E}}^{(+)}_{A}\left(\mathbf{r}\right) = \sqrt{\frac{\hbar \omega_0}{2\epsilon_0 c T}} \sum_{n=0}^{\infty} \hat{a}_n u_n \left(\mathbf{r}\right)
\end{equation}
with $u_n\left(\mathbf{r}\right)$ an orthonormal mode basis and $\hat{a}_n$ the associated modal annihilation operators.

For a beam displaced in the transverse plane (along $x$ for instance), and in the case where the mode basis $u_i\left(\mathbf{r} \right)$ is taken to be the Hermite-Gauss mode basis (also along $x$), we can write
\begin{align}
\hat{a}_0 =& \langle\hat{a}_0\rangle + \delta \hat{a}_0 =  \sqrt{N} + \delta \hat{a}_0\\
\hat{a}_1 =&\langle\hat{a}_1\rangle + \delta \hat{a}_1 = \frac{d}{w_0}\sqrt{N} + \delta \hat{a}_1 \\
\hat{a}_{i}=&\langle\hat{a}_i\rangle + \delta \hat{a}_i = o(d^2) + \delta \hat{a}_i, \,\, \forall \, i>1
\end{align}
where $d$ is the displacement of the beam.

The mode basis which is ``demultiplexed" by the MPLC system is slightly different from the Hermite-Gauss mode basis. The differences are revealed through the cross-talk phenomenon. We define the mode basis demultiplexed by the MPLC system as $v_i\left(\mathbf{r}\right)$ and $\hat{b}_n$ as the associated modal annihilation operators. We also define the cross-talk coeficients as
\begin{equation}
p^i_j = \int u_i^*\left(\mathbf{r}\right) v_j\left(\mathbf{r}\right) d\mathbf{r}.
\end{equation}
The demultiplexing operation does not introduce decoherence between the different modes which allows the use of a coherent description.  We also make the assumption that the demultiplexing operation is unitary in the sense that it does not introduce losses. We note that in the experiment we perform, the method of cross talk measurement only gives us access to $\left|p_j^i\right|$.
In a reasonable approximation considering the experimental values of cross-talk, we can write $p_i^i\simeq 1$ and:
\begin{align}
v_0 =& u_0 + p_1^0 u_1 + p_2^0 u_2 + ... \\
v_1 =& p_0^1 u_0 + u_1 + p_2^1 u_2 + ... 
\end{align}
At the input of the multiplexer, we express the incident field $\hat{\mathcal{E}}_{in}\left(\mathbf{r}\right) = \hat{\mathcal{E}}_A\left(\mathbf{r}\right) $ as
\begin{equation}
\hat{\mathcal{E}}^{(+)}_{in}\left(\mathbf{r}\right) = \sqrt{\frac{\hbar \omega_0}{2\epsilon_0 c T}} \sum_{n=0}^{\infty} \hat{b}_n v_n \left(\mathbf{r}\right).
\end{equation}
After the multiplexer and before the single mode fibers, we can describe the different spatially separated beams as:
\begin{align}
\hat{\mathcal{E}}_{out,i}^{(+)}\left(\mathbf{r}\right) &= \sqrt{\frac{\hbar \omega_0}{2\epsilon_0 c T}} \sum_{n=0}^{\infty} \hat{c}_{n,i} v^i_n \left(\mathbf{r}\right)  \\
& \text{ with } \hat{c}_{i,i} = \hat{b}_i \text{ and } \forall j \neq i, \langle\hat{c}_{j,i}\rangle=0,\nonumber
\end{align}
$v^i_n\left(\mathbf{r}\right)$ being a mode basis centered on the $i^{\text{th}}$ single mode fiber and $v^i_i\left(\mathbf{r}\right)$ corresponding to the mode of the fiber.

Using these notations, we can express the intensity measured after the demultiplexer on a given channel as:
\begin{equation}
\hat{I}_{i} = \int\hat{\mathcal{E}}_{out,i}^{(+)\dagger}\left(\mathbf{r}\right) \hat{\mathcal{E}}_{out,i}^{(+)}\left(\mathbf{r}\right) d\mathbf{r} = \frac{\hbar \omega_0}{2\epsilon_0 c T} \sum_{i=0}^{\infty}\hat{c}_{n,i}^{\dagger}\hat{c}_{ni} 
\end{equation}
The mean value of $\hat{I}_{i}$ is
\begin{equation}
\langle\hat{I}_{i}\rangle = \frac{\hbar \omega_0}{2\epsilon_0 c T} \langle \hat{c}_{n,i}^{\dagger}\hat{c}_{ni} \rangle = \frac{\hbar \omega_0}{2\epsilon_0 c T} \langle \hat{b}_{i}^{\dagger}\hat{b}_{i} \rangle.
\end{equation}
At small displacements, we are most interested in $\hat{I}_1$ and must thus express $\hat{b}^{\dagger}_1 \hat{b}_1$. We do so at first order in $d$, by only considering the contributions of $\hat{a}_0$ and $\hat{a}_1$:
\begin{align}
\hat{b}^{\dagger}_1 \hat{b}_1 =& \langle \hat{a}_1\rangle^2 + 2 p_0^1\langle \hat{a}_0\rangle \langle \hat{a}_1\rangle + (p_0^1)^2 \langle \hat{a}_0\rangle^2  \\ 
&+ \left( \langle \hat{a}_1\rangle + p_0^1 \langle \hat{a}_0\rangle \right) \left[\delta \hat{a}_1 + \delta \hat{a}_1^{\dagger} + p_0^1\left( \delta \hat{a}_0 + \delta \hat{a}_0^{\dagger} \right)\right] \nonumber \\
&+ \delta \hat{a}_1^{\dagger}\delta \hat{a}_1 + p_0^1 \delta \hat{a}_1^{\dagger}\delta \hat{a}_0 + p_0^1 \delta \hat{a}_0^{\dagger}\delta \hat{a}_1 + (p_0^1)^2\delta \hat{a}_0^{\dagger}\delta \hat{a}_0. \nonumber
\end{align}
From this expression, we can write that the intensity measured on mode $v_1$ is:
\begin{equation}
\langle\hat{I}_{1}\rangle = \frac{\hbar \omega_0}{2\epsilon_0 c T} N \left(\frac{d}{w_0} + p_0^1\right)^2
\end{equation}
and
\begin{equation}
\delta \hat{I}_{1} = \frac{\hbar \omega_0}{2\epsilon_0 c T} \sqrt{N} \left(\frac{d}{w_0}+p_0^1\right) \left(\delta \hat{X}_1 + p_0^1 \delta \hat{X}_0\right).
\end{equation}
For a coherent state at shot noise, $\langle\delta \hat{X}_0^2\rangle=\langle\delta \hat{X}_1^2\rangle=1$ and we can write
\begin{align}
\langle\delta \hat{I}_{1}^2\rangle =& \left(\frac{\hbar \omega_0}{2\epsilon_0 c T}\right)^2 \times N \times \left(\frac{d}{w_0}+p_0^1\right)^2 \\
& \times \left[1 + (p_0^1)^2  + p_0^1 \left(\langle \delta\hat{X}_0 \delta\hat{X}_1\rangle + \langle \delta\hat{X}_1 \delta\hat{X}_0\rangle \right) \right]. \nonumber
\end{align}

Finally, we give the expression of the signal to noise ratio:
\begin{equation}
\sqrt{\frac{\langle\hat{I}_1\rangle^2}{\langle\delta \hat{I}_1^2\rangle}} = \frac{\sqrt{N} \left(\frac{d}{w_0} + p_0^1 \right)} {\sqrt{1 + \left(p_0^1\right)^2 + p_0^1\left(\langle \delta\hat{X}_0 \delta\hat{X}_1\rangle + \langle \delta\hat{X}_1 \delta\hat{X}_0\rangle\right)}}
\label{eq:sensibilite}
\end{equation}

In the case where $p_0^1=0$ and for a signal to noise ratio equal to one, we find again that
\begin{equation}
\boxed{d = \frac{w_0}{\sqrt{N}}}
\end{equation}
which is consistent with the results of \cite{Delaubert08,Tsang16}.

If we now make the assumption that the field we consider is coherent, we have $\langle \delta\hat{X}_0 \delta\hat{X}_1\rangle=\langle \delta\hat{X}_1 \delta\hat{X}_0\rangle=0$. In the case where $p_0^1 \ll 1$, we can rewrite equation \ref{eq:sensibilite} as
\begin{equation}
1=\sqrt{\frac{\langle\hat{I}_1\rangle^2}{\langle\delta \hat{I}_1^2\rangle}} \simeq \sqrt{N} \left(\frac{d}{w_0} + p_0^1 \right)\left(1-\left(p_0^1\right)^2/2\right)
\end{equation}
 which at first order simplifies to 
\begin{equation}
\boxed{1=\sqrt{\frac{\langle\hat{I}_1\rangle^2}{\langle\delta \hat{I}_1^2\rangle}} = \sqrt{N} \left(\frac{d}{w_0} + p_0^1 \right).}
\end{equation}
The cross-talk coefficient thus appears as an offset term.

\bibliography{apssamp_ol}

\end{document}